\documentclass[preprint,showpacs,showkeys,preprintnumbers,amsmath,amssymb]{revtex4-1}
\usepackage{mathrsfs}
\usepackage{graphicx}
\usepackage{dcolumn}
\usepackage{bm}

\begin{document}

\title{From nonholonomic quantum constraint to canonical variables of photons I: true intrinsic degree of freedom}

\author{Chun-Fang Li\footnote{Corresponding author. Email address: cfli@shu.edu.cn} and Yun-Long Zhang\footnote{Email address: ylzhang@i.shu.edu.cn}}

\affiliation{Department of Physics, Shanghai University, 99 Shangda Road, 200444
Shanghai, China}

\date{\today}

\begin{abstract}

We report that the true intrinsic degree of freedom of the photon is neither the polarization nor the spin. It describes a local property in momentum space and is represented in the local representation by the Pauli matrices. This result is achieved by treating the transversality condition on the vector wavefunction as a nonholonomic quantum constraint.
We find that the quantum constraint makes it possible to generalize the Stokes parameters to characterize the polarization of a general state. Unexpectedly, the generalized Stokes parameters are specified in a momentum-space local reference system that is fixed by another degree of freedom, called Stratton vector.
Only constant Stokes parameters in one particular local reference system can convey the intrinsic degree of freedom of the photon.
We show that the optical rotation is one of such processes that change the Stratton vector with the intrinsic quantum number remaining fixed. Changing the Stratton vector of the eigenstate of the helicity will give rise to a Berry's phase.

\end{abstract}


\maketitle


\section{Introduction}

The momentum-space wavefunction of free photons, apart from satisfying Schr\"{o}dinger's dynamical equation, obeys the transversality condition, which is expressed in terms of the time-independent wavefunction $\mathbf{f} (\mathbf{k})$ as follows \cite{AB, CDG},
\begin{equation}\label{TC}
    \mathbf{k} \cdot \mathbf{f}=0,
\end{equation}
where $\mathbf k$ is the wavevector. This condition makes it hard to develop the formalism of the quantum mechanics for the photon.
Firstly, it makes the spin not independent of the orbital angular momentum \cite{AB, CDG, EN, Barn, LL}. In particular, the total angular momentum cannot be strictly split into helicity-dependent spin and helicity-independent orbital parts \cite{BA, Li09-1, Li-WY}.
If we insist \cite{Merz} that a degree of freedom in quantum mechanics, either extrinsic or intrinsic, should be an independent physical variable the same as in classical mechanics \cite{Gold}, the spin cannot be the intrinsic degree of freedom of the photon \cite{Birula} though it is misleadingly called that frequently in the literature \cite{MMP, Naga, Marr, SYM, BRNZ}.
Secondly, the notion of polarization that classically refers to the direction of vibration of the transverse electric vector is also usually regarded as the intrinsic degree of freedom \cite{BRNZ, LZ, Kwiat, BAOA, BSCB}. But because the vector wavefunction $\mathbf f$ is substantially the electric vector of the plane-wave constituent at the momentum $\mathbf k$ as is indicated by the integral expression of the electric vector in position space \cite{AB, CDG},
\begin{equation*}
    \mathfrak{E} (\mathbf{x},t)
   =\frac{1}{(2 \pi)^{3/2}}
    \int \bigg( \frac{\hbar \omega}{2\varepsilon_0} \bigg)^{1/2}
    \mathbf{f}(\mathbf{k}) e^{i (\mathbf{k} \cdot \mathbf{x}-\omega t)} d^3 k
   +\text{c.c.},
\end{equation*}
where $\omega= c k$ and $k=|\mathbf{k}|$,
it follows from the transversality condition (\ref{TC}) that the polarization is not independent of the momentum and thus cannot be a degree of freedom at all. In words, the true intrinsic degree of freedom of the photon is not as clear as might be thought.

It is known \cite{CDG, Merz} that a wavefunction of multiple components describes the intrinsic degree of freedom of quantum particles if all its components are independent. Here the three Cartesian components of the vector wavefunction $\mathbf f$ are not independent. They are connected with one another by Eq. (\ref{TC}). So it is understandable that they do not describe the intrinsic degree of freedom of the photon.
It is also known \cite{Gold} that if a system of particles in classical mechanics is free from any constraints, the Cartesian coordinates of the particles in the laboratory reference system, independent of one another, can serve as degrees of freedom. But if there exist constraints, not all the Cartesian coordinates can serve as degrees of freedom. In that case, the constraints are used to introduce new independent variables, called generalized coordinates, as the degrees of freedom, which usually differ from the Cartesian coordinates. The purpose of this paper is to treat the transversality condition (\ref{TC}) as a quantum constraint to explore the true intrinsic degree of freedom of the photon.

It is noted that Eq. (\ref{TC}) is only a constraint on a single photon because the vector wavefunction is about a single photon.
But the same as the classical constraints are used to introduce the generalized coordinates, this constraint can also be used to introduce a new wavefunction that consists of new two independent functions, $f_1 (\mathbf{k})$ and $f_2 (\mathbf{k})$, in terms of which the vector wavefunction is expressed by equation of the form
\begin{equation}\label{TE}
    \mathbf{f}=\mathbf{f} (f_1, f_2),
\end{equation}
containing the constraint (\ref{TC}) in it implicitly. However, expressed in terms of the momentum, the constraint (\ref{TC}) is nonholonomic in the language of classical mechanics. We will see that the new two-component wavefunction introduced from such a constraint is not defined in the laboratory reference system in which the vector wavefunction is defined, a phenomenon that is analogous to the fact that the generalized coordinates in classical mechanics are not the Cartesian coordinates. The consequences of this result are twofold.

In the first place, the intrinsic degree of freedom of the photon is unexpectedly not a physical quantity of the photon with respect to the laboratory reference system. In the second place, even though the quantum constraint (\ref{TC}) is a constraint on a single photon, the canonical position that is conjugate to the momentum through the Fourier integral of the new two-component wavefunction is not the position of the photon in the laboratory reference system.
In present paper we are concerned only with the intrinsic degree of freedom. The canonical position will be discussed in a subsequent paper. The main results of this paper are summarized as follows.

As is known, the most convenient mathematical characterization for the polarization of a plane-wave state is the Stokes parameters (SPs) \cite{JR}, which are completely determined by its two-component Jones vector via the Pauli matrices \cite{Dama}.
By generalizing the Jones vector of a plane-wave state, a new two-component wavefunction is introduced in Section \ref{Intro-SP} from the quantum constraint (\ref{TC}). Related to the vector wavefunction by a quasi unitary transformation, it determines the SPs to characterize the polarization of a general state.
However, it is found in Section \ref{Local-SP} that the nonholonomic characteristic of the constraint (\ref{TC}) makes it necessary for a constant unit vector, called Stratton vector (SV) \cite{Stra}, to fix the SPs. So fixed SPs are exactly specified in the momentum-space local reference system (LRS) that is fixed by the same SV. As a result, associated with any LRS there is a local representation in which the SPs are represented by the same Pauli matrices.
Based on these results, it is discussed in Section \ref{IDOF} that the constant SPs in each local representation convey the intrinsic degree of freedom of the photon in the corresponding LRS.
Since the SU(2) commutation relation of the Pauli matrices guarantees that so identified intrinsic degree of freedom is canonically quantized, the SV to fix the LRS appears as another degree of freedom. It combines with the intrinsic quantum number to jointly determine the polarization state of the photon.
One of the observable effects of the SV degree of freedom is analyzed in Section \ref{Effects}. It is shown that the optical rotation \cite{Dama} is such a physical process that changes the SV degree of freedom with the intrinsic quantum number remaining fixed.
Section \ref{Remarks} concludes the paper with remarks.

\section{Quantum constraint and generalization of SPs}\label{Intro-SP}

The quantum constraint (\ref{TC}) means that at any momentum there exist two mutually-orthogonal base vectors in terms of which the vector wavefunction at that momentum can be expanded.
For a plane-wave state, it is always allowed to assume that the momentum goes along a fixed direction, say the $z$ axis.
But for a general state, it is impossible to assume that the momenta of all its plane-wave constituents go along a fixed direction. So we consider at any momentum two mutually-perpendicular real unit vectors $\mathbf u$ and $\mathbf v$ that form with the unit wavevector
$\mathbf{w}=\frac{\mathbf{k}}{k}$, at the same momentum,
a right-handed Cartesian system satisfying
\begin{equation}\label{triad}
    \mathbf{u} \times \mathbf{v} =\mathbf{w}, \quad
    \mathbf{v} \times \mathbf{w} =\mathbf{u}, \quad
    \mathbf{w} \times \mathbf{u} =\mathbf{v}.
\end{equation}
For later comparison with the Pauli matrices for the spin of the electron, we choose, instead of linearly-polarized base vectors, circularly-polarized base vectors
$\mathbf{c}_{+} =\frac{1}{\sqrt 2} (\mathbf{u} +i \mathbf{v})$
and
$\mathbf{c}_{-} =\frac{1}{\sqrt 2} (\mathbf{u} -i \mathbf{v})$
to expand the vector wavefunction as
\begin{equation*}
    \mathbf{f}(\mathbf{k})=\mathbf{c}_{+} f_{+} (\mathbf{k})
                          +\mathbf{c}_{-} f_{-} (\mathbf{k}).
\end{equation*}
Since this expansion for the vector wavefunction automatically satisfies the constraint (\ref{TC}), the expansion coefficients $f_{+}$ and $f_{-}$ are the new two independent functions that we introduce. For a plane-wave state, they are the two components of the Jones vector \cite{Dama}.
We generalize the Jones vector by letting these two independent functions form a two-component entity
$\tilde{f}=\bigg(\begin{array}{c}
                   f_{+} \\
                   f_{-}
                 \end{array}
           \bigg)
$
in terms of which the above expansion is rewritten as \cite{Li09-1, Li-WY, Li}
\begin{equation}\label{QUT1}
    \mathbf{f}(\mathbf{k})=\varpi \tilde{f}(\mathbf{k}),
\end{equation}
where vectors of three Cartesian components such as $\mathbf{c}_{+}$ and $\mathbf{c}_{-}$ are expressed as column matrices and
$
\varpi=(\begin{array}{cc}
              \mathbf{c}_{+} & \mathbf{c}_{-} \\
            \end{array}
       )
$
is a 3-by-2 matrix. Eq. (\ref{QUT1}) is what we give for the transformation (\ref{TE}). According to Eqs. (\ref{triad}), we have
\begin{equation}\label{unitarity1}
    \varpi^{\dag} \varpi =I_2,
\end{equation}
from which it follows that
\begin{equation*}
    \mathbf{f}^\dag (\mathbf{k}) \mathbf{f} (\mathbf{k})
   =\tilde{f}^\dag (\mathbf{k}) \tilde{f} (\mathbf{k}),
\end{equation*}
where the superscript $\dag$ stands for the complex transpose and $I_2$ is the 2-by-2 unit matrix. This shows that the two-component entity also serves as a momentum-space wavefunction, referred to as the Jones wavefunction.

With the Jones wavefunction, we are in a position to generalize the SPs.
Considering that the two components of the Jones wavefunction are independent of each other, we split the unit Jones wavefunction $\tilde{a}$ that satisfies the normalization
\begin{equation}\label{NC}
    \tilde{a}^\dag \tilde{a} =1
\end{equation}
by writing the Jones wavefunction as
\begin{equation*}
    \tilde{f}(\mathbf{k})=\tilde{a} f(\mathbf{k}),
\end{equation*}
where $f(\mathbf{k})$ is any physically allowed complex function.
Substituting it into Eq. (\ref{QUT1}), we have
$\mathbf{f}(\mathbf{k})=\mathbf{a} f(\mathbf{k})$,
where
\begin{equation}\label{QUT1-a}
    \mathbf{a}=\varpi \tilde{a}
\end{equation}
is the unit vector wavefunction satisfying
\begin{equation}\label{unitarity}
    \mathbf{a}^\dag \mathbf{a} =\tilde{a}^\dag \tilde{a}
\end{equation}
by virtue of Eq. (\ref{unitarity1}). The unit vector wavefunction, which is known in the literature \cite{AB} as the polarization vector, describes the polarization of the photon state $\mathbf f$.
With the polarization vector, the quantum constraint (\ref{TC}) is equally expressed as
\begin{equation}\label{QC}
    \mathbf{k} \cdot \mathbf{a}=0,
\end{equation}
which explicitly states that the polarization vector is dependent on the momentum.
It is noted that the matrix $\varpi$ in Eq. (\ref{QUT1}) or (\ref{QUT1-a}) arises only from the quantum constraint (\ref{TC}) or (\ref{QC}). It has nothing to do with the concrete form of the vector wavefunction or of the polarization vector.
Multiplying both sides of Eq. (\ref{QUT1-a}) by $\varpi^{\dag}$ from the left and considering Eq. (\ref{unitarity1}), we have
\begin{equation}\label{QUT2-a}
    \tilde{a}=\varpi^{\dag} \mathbf{a}.
\end{equation}
The same as the SPs of a plane-wave state are determined by its unit Jones vector, the SPs of a general state $\mathbf f$ at any momentum are stipulated to be determined by the unit Jones wavefunction at the same momentum in the following form,
\begin{equation}\label{SP}
    \varsigma_i =\tilde{a}^\dag \hat{\sigma}_i \tilde{a}, \quad i=1,2,3,
\end{equation}
where the Pauli matrices
\begin{equation}\label{PM}
    \hat{\sigma}_1=\bigg(\begin{array}{cc}
                           0 & 1 \\
                           1 & 0
                         \end{array}
                   \bigg), \quad
    \hat{\sigma}_2=\bigg(\begin{array}{cc}
                           0 & -i \\
                           i &  0
                         \end{array}
                   \bigg), \quad
    \hat{\sigma}_3=\bigg(\begin{array}{cc}
                           1 &  0 \\
                           0 & -1
                         \end{array}
                   \bigg)
\end{equation}
are the same as those for the spin of the electron.

The matrix $\varpi$ in Eqs. (\ref{QUT1-a}) and (\ref{QUT2-a}) performs a quasi unitary transformation in the following sense.
On one hand, Eq. (\ref{QUT1-a}) says that the matrix $\varpi$ acts on a unit Jones wavefunction to yield a unit vector wavefunction.
On the other hand, Eq. (\ref{QUT2-a}) says that the matrix $\varpi^{\dag}$ acts on a unit vector wavefunction to yield a unit Jones wavefunction.
Substituting Eq. (\ref{QUT2-a}) into the right-handed side of Eq. (\ref{unitarity}) and considering the arbitrariness of $\mathbf a$, we get
\begin{equation}\label{unitarity2}
    \varpi \varpi^{\dag} =I_3,
\end{equation}
where $I_3$ is the 3-by-3 unit matrix. Eqs. (\ref{unitarity1}) and (\ref{unitarity2}) express the quasi unitarity \cite{Golub} of the transformation matrix $\varpi$. $\varpi^{\dag}$ is the Moore-Penrose pseudo inverse of $\varpi$, and vice versa.
The quasi unitarity of $\varpi$ guarantees that to each polarization state $\mathbf a$ there corresponds a unique unit Jones wavefunction via Eq. (\ref{QUT2-a}) and hence a unique set of SPs via Eq. (\ref{SP}).
This demonstrates that the quantum constraint makes it possible to generalize the SPs to characterize the polarization of a general state.

Now that the two components of the unit Jones wavefunction are independent of each other, it seems that the SPs determined by the unit Jones wavefunction would convey the intrinsic degree of freedom of the photon. Unfortunately, things are not so simple.
The key point is that the quantum constraint (\ref{QC}) cannot solely fix the transverse axes, $\mathbf u$ and $\mathbf v$, of the Cartesian system $\mathbf{uvw}$, or the quasi unitary matrix $\varpi$, up to a rotation about $\mathbf k$ \cite{Mess, MW}. That is to say, the constraint (\ref{QC}) cannot solely fix the unit Jones wavefunction (\ref{QUT2-a}) and the SPs (\ref{SP}). This reflects the nonholonomic characteristic of the constraint (\ref{QC}).
To know how the SPs can convey the intrinsic degree of freedom, it is necessary to make clear how the SPs can be fixed and what so fixed SPs mean.

\section{Locality of SPs in momentum space}\label{Local-SP}

\subsection{Introduction of SV}

It was shown a long time ago by Stratton \cite{Stra} and later on by others \cite{GW, PA, DP} that the transverse axes of the Cartesian system $\mathbf{uvw}$ at any momentum can be fixed consistently by introducing an arbitrary real constant unit vector $\mathbf I$, the so-called SV, in the following way,
\begin{equation}\label{axes}
    \mathbf{u} =\mathbf{v}               \times \frac{\mathbf k}{k},          \quad
    \mathbf{v} =\frac{\mathbf{I} \times \mathbf{k}}{|\mathbf{I} \times \mathbf{k}|}.
\end{equation}
It is easy to check that so fixed unit vectors satisfy Eqs. (\ref{triad}) irrespective of the SV.
By this it is meant that any particular SV is able to fully fix the quasi unitary matrix $\varpi$ and therefore the SPs via Eqs. (\ref{QUT2-a})-(\ref{PM}).
Mathematically, any particular SV can fix through the quasi unitary transformation (\ref{QUT2-a}) a two-component representation in which the SPs (\ref{SP}) are represented by the same Pauli matrices (\ref{PM}). As a consequence, a polarization state generally has different SPs in different two-component representation.
Observing that the SV actually fixes through Eqs. (\ref{axes}) the LRS $\mathbf{uvw}$ in momentum space, corresponding to each two-component representation there is a particular LRS.
The SPs in each two-component representation are expected to be specified in its corresponding LRS.
For this reason, the two-component representation is said to be the local representation for the corresponding LRS.
To find out how the SPs are specified in the LRS, let us first analyze how the SPs change with the local representation.

Consider a change of the local representation from $\mathbf I$ to a different one, say $\mathbf{I}'$. In this case, the unit Jones wavefunction of the same polarization state $\mathbf a$ is given by
\begin{equation}\label{QUT2'}
    \tilde{a}'=\varpi'^{\dag} \mathbf{a},
\end{equation}
where
$\varpi'=(\begin{array}{cc}
            \mathbf{c}'_{+} & \mathbf{c}'_{-}
          \end{array}
         ),
$
$\mathbf{c}'_{+} =\frac{1}{\sqrt 2} (\mathbf{u}' +i \mathbf{v}')$,
$\mathbf{c}'_{-} =\frac{1}{\sqrt 2} (\mathbf{u}' -i \mathbf{v}')$,
and
\begin{equation*}
    \mathbf{u}' =\mathbf{v}' \times \frac{\mathbf k}{k}, \hspace{5pt}
    \mathbf{v}' =\frac{\mathbf{I}' \times \mathbf{k}}{|\mathbf{I}' \times \mathbf{k}|}.
\end{equation*}
As mentioned above, the transverse axes
$\mathbf{u}'$ and $\mathbf{v}'$
of the primed LRS $\mathbf{I}'$ are related to the transverse axes $\mathbf{u}$ and $\mathbf{v}$ of the unprimed one $\mathbf I$ by a rotation about $\mathbf k$.
Denoted by $\Phi(\mathbf{k}; \mathbf{I}', \mathbf{I})$, the rotation angle is determined by
\begin{equation}\label{Varpi'}
    \varpi'=\exp [-i (\hat{\mathbf \Sigma} \cdot \mathbf{w}) \Phi] \varpi
\end{equation}
or by
\begin{equation}\label{varpi'}
    \varpi'=\varpi \exp \left(-i \hat{\sigma}_3 \Phi \right),
\end{equation}
where
$(\hat{\Sigma}_k)_{ij} =-i \epsilon_{ijk}$
with $\epsilon_{ijk}$ the Levi-Civit\'{a} pseudotensor.
Substituting Eq. (\ref{varpi'}) into Eq. (\ref{QUT2'}) and noticing Eq. (\ref{QUT2-a}), we have
\begin{equation}\label{GT-a}
    \tilde{a}'=\exp \left(i \hat{\sigma}_3 \Phi \right) \tilde{a}.
\end{equation}
This is the transformation law of the unit Jones wavefunction under the change of the local representation.
It is emphasized that the transformation function $\Phi$ always depends on the momentum, unless $\mathbf{I}' =-\mathbf{I}$. In that case, we have $\Phi=\pi$.

According to the definition (\ref{SP}), the SPs in the primed local representation $\mathbf{I}'$ are given by
\begin{equation}\label{SP'}
    \varsigma'_i =\tilde{a}'^\dag \hat{\sigma}_i \tilde{a}'.
\end{equation}
Substituting Eq. (\ref{GT-a}) into Eq. (\ref{SP'}) and using Eq. (\ref{SP}), we get
\begin{subequations}\label{GT-SP}
\begin{align}
  \varsigma'_1 & = \varsigma_1 \cos 2\Phi+\varsigma_2 \sin 2\Phi, \label{SP1} \\
  \varsigma'_2 & =-\varsigma_1 \sin 2\Phi+\varsigma_2 \cos 2\Phi, \label{SP2} \\
  \varsigma'_3 & = \varsigma_3,                                   \label{SP3}
\end{align}
\end{subequations}
which constitute the transformation law of the SPs under the change of the local representation.
From this result it is concluded that the \emph{nonholonomic characteristic of the constraint makes it necessary for the SV to fix the SPs}.
Due to the dependence of the transformation function $\Phi$ on the momentum, the SPs in an arbitrary local representation generally depend on the momentum and hence cannot convey the intrinsic degree of freedom of the photon.

\subsection{Specification of SPs in LRS}

The SPs of a plane wave, which constitute a unit vector, are generally specified on the surface of the Poincar\'{e} sphere \cite{Dama}.
Here we will make use of the transformation law (\ref{GT-SP}) to show that the SPs in each local representation are specified in its corresponding LRS in the sense that they denote the components of a unit vector along the Cartesian axes of that LRS.

On one hand, the transformation equations (\ref{GT-SP}), resulting from a rotation of the LRS about $\mathbf k$, do reveal that the SPs in each local representation are the Cartesian components of a unit vector.
In particular, Eq. (\ref{SP3}) shows that the third parameter is the component along the direction of the momentum, the longitudinal component, and Eqs. (\ref{SP1}) and (\ref{SP2}) show that the first two parameters are the mutually-perpendicular transverse components.
On the other hand, corresponding to the rotation (\ref{Varpi'}) of the transverse axes of the LRS by an angle of $\Phi$, the transverse SPs are rotated by an angle of $-2\Phi$.
The only possible explanation of this relation is that the SPs in different local representation do not stand for the same unit vector.
Noticing that the transformation of the local representation has a one-to-one correspondence with the rotation of the LRS about $\mathbf k$, it is justified to postulate that the SPs in each local representation denote the components of a unit vector along the Cartesian axes of the corresponding LRS.
Specifically, the SPs (\ref{SP}) in the unprimed local representation constitute the following unit vector in the unprimed LRS,
\begin{equation}\label{UV}
    \boldsymbol{\varsigma}
   =\varsigma_1 \mathbf{u} +\varsigma_2 \mathbf{v} +\varsigma_3 \mathbf{w}.
\end{equation}
Likewise, the SPs (\ref{SP'}) in the primed local representation constitute the following unit vector in the primed LRS,
\begin{equation}\label{UV'}
    \boldsymbol{\varsigma}'
   =\varsigma'_1 \mathbf{u}' +\varsigma'_2 \mathbf{v}' +\varsigma'_3 \mathbf{w}.
\end{equation}
Substituting Eqs. (\ref{GT-SP}) into Eq. (\ref{UV'}) and taking Eqs. (\ref{varpi'}) and (\ref{UV}) into account, we find
\begin{equation}\label{GT-UV}
    \boldsymbol{\varsigma}'
   =\exp{[i(\hat{\mathbf \Sigma} \cdot \mathbf{w})\Phi]} \boldsymbol{\varsigma},
\end{equation}
which confirms our assertion that the unit vectors (\ref{UV}) and (\ref{UV'}) are not the same. The meaning of this relation is explained in detail below with the transverse SPs at one particular momentum.

Eqs. (\ref{SP1}) and (\ref{SP2}) indicate that if the SPs $\varsigma'_1$ and $\varsigma'_2$ were specified in the unprimed LRS in which the SPs $\varsigma_1$ and $\varsigma_2$ are specified, they would constitute a transverse component,
$\boldsymbol{\varsigma}''_\perp=\varsigma'_1 \mathbf{u}+\varsigma'_2\mathbf{v}$,
that is the result of the rotation of the transverse component
$\boldsymbol{\varsigma}_\perp=\varsigma_1 \mathbf{u} +\varsigma_2 \mathbf{v}$
by an angle of $-2 \Phi$ as is displayed in Fig. 1(a).
But as just pointed out, they should be specified in the primed LRS. Along with the unprimed LRS being rotated to the primed LRS by an angle of $\Phi$, the transverse component $\boldsymbol{\varsigma}''_\perp$ is rotated to
$\boldsymbol{\varsigma}'_\perp=\varsigma'_1 \mathbf{u}'+\varsigma'_2 \mathbf{v}'$ as is displayed in Fig. 1(b).
Consequently, $\boldsymbol{\varsigma}'_\perp$ is equal to the result of the rotation of $\boldsymbol{\varsigma}_\perp$ by an angle of $-\Phi$. This is what Eq. (\ref{GT-UV}) means.
\begin{figure}[tb]
\centerline{\includegraphics[width=11cm]{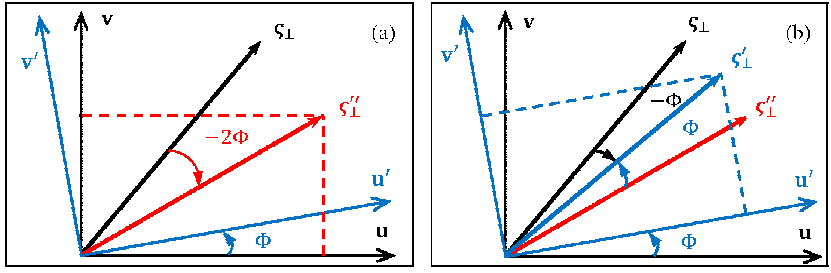}}
\caption{(a) $\boldsymbol{\varsigma}''_\perp$ is the result of the rotation of $\boldsymbol{\varsigma}_\perp$ by an angle of $-2 \Phi$. (b) $\boldsymbol{\varsigma}''_\perp$ is rotated to $\boldsymbol{\varsigma}'_\perp$ along with the LRS $\mathbf I$ being rotated to the LRS $\mathbf{I}'$ by an angle of $\Phi$.}
\end{figure}

To conclude this section we summarize our main results. The SPs (\ref{SP}) in each local representation are specified in its corresponding LRS according to Eq. (\ref{UV}). Mathematically, the unit Jones wavefunction (\ref{QUT2-a}) in each local representation is defined in its corresponding LRS.

\section{Identification of intrinsic degree of freedom}\label{IDOF}

\subsection{From SPs to intrinsic degree of freedom}

It is now clear from Eqs. (\ref{UV}) and (\ref{UV'}) that a polarization state can be characterized in different local representation by different SPs.
Eqs. (\ref{GT-SP}) are the transformation law of the SPs from the local representation $\mathbf I$ to the local representation $\mathbf{I}'$, corresponding to the transformation law (\ref{GT-a}) of the unit Jones wavefunction. In that sense, different local representations are equivalent.
Now that the SPs in each local representation are specified in its corresponding LRS, it is no wonder why the SPs of a given polarization state in an arbitrary LRS generally depend on the momentum.
However, from the fact that the SPs (\ref{SP}) in any local representation are represented by the same Pauli matrices (\ref{PM}) we will see that the intrinsic degree of freedom of the photon is one of its quantum property with respect to any particular LRS.

To show this, it is emphasized that the transversality condition (\ref{QC}) makes the polarization vector depend on the momentum. Because the base vectors composing the quasi unitary matrix $\varpi$ satisfy the transversality condition, there must be polarization states the unit Jones wavefunctions of which in some LRS are independent of the momentum, giving constant SPs.
Only such constant SPs can convey the intrinsic degree of freedom of the photon. The peculiarity here is that constant SPs in one LRS are generally not constant in others. This lies again in the dependence of the transformation function $\Phi$ on the momentum.
Consider a polarization state the unit Jones wavefunction of which in the LRS $\mathbf I$ is a constant function $\tilde \alpha$, giving constant SPs
$n_i=\tilde{\alpha}^\dag \hat{\sigma}_i \tilde{\alpha}$.
If the transverse SPs $n_1$ and $n_2$ do not vanish simultaneously, it follows from Eqs. (\ref{SP1}) and (\ref{SP2}) that the transverse SPs of the same polarization state in any other LRS $\mathbf{I}'$ are no longer constant, unless $\mathbf{I}'=-\mathbf{I}$.
On the other hand, since the SPs in any local representation are represented by the same Pauli matrices (\ref{PM}), different constant unit Jones wavefunction in one particular local representation gives different constant SPs, conveying different value of the intrinsic degree of freedom with respect to the same LRS.
By this it is meant that \emph{the intrinsic degree of freedom of the photon is a quantity of the photon with respect to any particular LRS} and is represented in the associated local representation by the Pauli matrices (\ref{PM}).
According to Eq. (\ref{UV}), it appears in that LRS as a ``constant'' unit vector,
$\mathbf{n}=n_1 \mathbf{u}+n_2 \mathbf{v}+n_3 \mathbf{w}$.
It is important to note that so identified intrinsic degree of freedom is canonically quantized as is guaranteed by the SU(2) commutation relation of the Pauli matrices,
\begin{equation}\label{CCR}
    [\hat{\sigma}_i, \hat{\sigma}_j]=2i \sum_k \epsilon_{ijk} \hat{\sigma}_k.
\end{equation}
Any intrinsic quantum number following from this commutation relation should be interpreted as a quantum variable with respect to some LRS.
In a word, the intrinsic degree of freedom of the photon is a local property of the photon in momentum space and is represented in the local representation by the Pauli matrices.

It can be inferred from the above discussions that in different local representations, the same value of the intrinsic degree of freedom described by the same constant unit Jones wavefunction does not stand for the same state of polarization. In this sense, different local representations are not equivalent.
The SV is not simply a mathematical tool that fixes the LRS to specify the SPs. In quantum mechanics, it also \emph{appears as another independent degree of freedom}.
Only combined with it can the intrinsic degree of freedom determine the polarization state of the photon, which is expressed in terms of the quasi unitary matrix $\varpi$ and the constant unit Jones wavefunction $\tilde \alpha$ as follows,
\begin{equation}\label{QUT1-al}
    \mathbf{a}_\mathbf{I}=\varpi \tilde{\alpha}.
\end{equation}

\subsection{Connection with spin}

To further illustrate the locality of the intrinsic degree of freedom in momentum space, let us compare it with the spin. As is known \cite{CDG, Li09-1}, the spin angular momentum of the photon in a state $\mathbf f$ is given by
$
\mathbf{S}=\frac{\int \mathbf{f}^\dag \hat{\mathbf S} \mathbf{f} d^3 k}
                {\int \mathbf{f}^\dag \mathbf{f} d^3 k}
$,
where
$\hat{\mathbf S}=\hbar \hat{\mathbf \Sigma}$
is the spin operator.
Represented by $\hat{\mathbf \Sigma}$ in the laboratory representation, the spin is supposed to be a vector quantity in the laboratory reference system.
If the constraint (\ref{TC}) were absent, the Cartesian components of the spin operator would satisfy the canonical commutation relations,
\begin{equation*}
    [\hat{S}_x, \hat{S}_y]=i \hbar \hat{S}_z, \quad
    [\hat{S}_y, \hat{S}_z]=i \hbar \hat{S}_x, \quad
    [\hat{S}_z, \hat{S}_x]=i \hbar \hat{S}_y.
\end{equation*}
Now that the vector wavefunction satisfies the condition (\ref{TC}), we have
\begin{equation*}
    \mathbf{f}^\dag (\hat{\mathbf \Sigma} \times \mathbf{w}) \mathbf{f}=0,
\end{equation*}
where we have used the relation
$\mathbf{f}^\dag \hat{\mathbf \Sigma} \mathbf{f}=-i \mathbf{f}^* \times \mathbf{f}$ \cite{CDG}.
Upon taking into account of the decomposition
$\hat{\mathbf \Sigma}=(\hat{\mathbf \Sigma}\cdot \mathbf{w}) \mathbf{w}
                     -(\hat{\mathbf \Sigma}\times \mathbf{w})\times \mathbf{w}$,
the spin operator reduces to
\begin{equation}\label{SO}
    \hat{\mathbf S}=\hbar (\hat{\mathbf \Sigma} \cdot \mathbf{w}) \mathbf{w},
\end{equation}
in consistency with the well-known conclusion \cite{MW, JR} that the spin of the photon lies entirely along the direction of the momentum. As a result, the spin cannot be canonically quantized \cite{EN},
\begin{equation*}
    [\hat{S}_x, \hat{S}_y]=[\hat{S}_y, \hat{S}_z]=[\hat{S}_z, \hat{S}_x]=0.
\end{equation*}
But the spin does have definite relation with the intrinsic degree of freedom.

To show this, we make use of Eqs. (\ref{SO}) and (\ref{QUT1}) and take Eq. (\ref{triad}) into account to rewrite the spin angular momentum in the local representation as
\begin{equation}\label{S}
    \mathbf{S}=\frac{\int \tilde{f}^\dag \hat{\mathbf s} \tilde{f} d^3 k}
                    {\int \tilde{f}^\dag \tilde{f} d^3 k},
\end{equation}
where
$
\hat{\mathbf s}=\hbar \hat{\sigma}_3 \mathbf{w}
$
is the spin operator and
$\hat{\sigma}_3=\varpi^\dag (\hat{\mathbf \Sigma}\cdot \mathbf{w}) \varpi$.
It is seen that the spin is essentially the longitudinal component of the intrinsic degree of freedom. That is to say, the helicity, the amplitude of the spin, is the longitudinal component of the intrinsic degree of freedom. It is noticed that the spin (\ref{S}) expressed in the local representation is invariant under the transformation (\ref{GT-a}). This just reflects the fact that the spin is a quantity in the laboratory reference system.

\section{Quantum mechanical effects of SV}\label{Effects}

\subsection{Description of an observable effect of SV}

We have deduced from the nonholonomic constraint (\ref{QC}) that the intrinsic degree of freedom of the photon is just a local quantity in momentum space. It is represented in the local representation by the Pauli matrices (\ref{PM}). The local representation is fixed by another degree of freedom, the SV. Only combined with the SV can the intrinsic degree of freedom determine the polarization state.
The recognition of the SV degree of freedom means that there should be two independent mechanisms to change the polarization state of a photon. One is to change the intrinsic degree of freedom with the SV remaining fixed. The other is to change the SV with the intrinsic degree of freedom remaining fixed.
The change of polarization state by use of polarizers \cite{Dama} belongs to the former. Let us explain the meaning of the latter, the newly identified mechanism, and discuss how to implement it.

To this end, we compare two photon states that are described by the same Jones wavefunction
\begin{equation*}
    \tilde{f}(\mathbf{k})=\tilde{\alpha} f(\mathbf{k})
\end{equation*}
in two different local representations $\mathbf I$ and $\mathbf{I}'$, where $\tilde{\alpha}$ is a constant unit Jones wavefunction. Of course, their intrinsic degrees of freedom take the same value.
The first state that is described in the local representation $\mathbf I$ has vector wavefunction
$\mathbf{f}_\mathbf{I}=\mathbf{a}_\mathbf{I} f(\mathbf{k})$,
where the polarization vector $\mathbf{a}_\mathbf{I}$ is given by Eq. (\ref{QUT1-al}).
Accordingly, the second state described in the local representation $\mathbf{I}'$ has vector wavefunction
$\mathbf{f}_{\mathbf{I}'} =\mathbf{a}_{\mathbf{I}'} f(\mathbf{k})$,
where
\begin{equation}\label{QUT1'-al}
    \mathbf{a}_{\mathbf{I}'}=\varpi' \tilde{\alpha}.
\end{equation}
Substituting Eq. (\ref{Varpi'}) into Eq. (\ref{QUT1'-al}) and noticing Eq. (\ref{QUT1-al}), we have
\begin{equation}\label{a'-a}
    \mathbf{a}_{\mathbf{I}'}
   =\exp [-i(\hat{\mathbf \Sigma} \cdot \mathbf{w}) \Phi] \mathbf{a}_\mathbf{I}.
\end{equation}
This shows that the state of polarization can indeed be changed by changing the SV with the intrinsic degree of freedom remaining fixed.

To show how such a change can be experimentally implemented, we simplify our discussions and assume that the two states are plane waves of the same momentum $\mathbf{k}_0$ so that their common Jones wavefunction takes the concrete form
$\tilde{f}_{\mathbf{k}_0}=\tilde{\alpha} \delta^3 (\mathbf{k}-\mathbf{k}_0)$.
In this case, the polarization vectors (\ref{QUT1-al}) and (\ref{QUT1'-al}) become
\begin{subequations}\label{a-k}
\begin{align}
  \mathbf{a}_{\mathbf{I}, \mathbf{k}_0} &=\varpi_0 \tilde{\alpha}, \label{ak0}\\
  \mathbf{a}_{\mathbf{I}',\mathbf{k}_0} &=\varpi'_0 \tilde{\alpha},\label{ak0'}
\end{align}
\end{subequations}
respectively, where the subscript 0 denotes the value taken at the momentum $\mathbf{k}_0$. Accordingly, Eq. (\ref{a'-a}) reduces to a rotation about the propagation direction,
\begin{equation}\label{a'-a-k}
    \mathbf{a}_{\mathbf{I}', \mathbf{k}_0}
   =\exp [-i(\hat{\mathbf \Sigma} \cdot \mathbf{w}_0) \phi]
    \mathbf{a}_{\mathbf{I}, \mathbf{k}_0}.
\end{equation}
It is noted that since a plane wave propagates in a fixed direction, the rotation angle $\phi$ here is no longer restricted by Eq. (\ref{Varpi'}) or (\ref{varpi'}).
As a matter of fact, it is seen from Eqs. (\ref{axes}) that Eq. (\ref{a'-a-k}) can be interpreted as rotating the SV of a plane wave about the propagation direction by the angle $\phi$. In other words, rotating the SV of a plane wave about the propagation direction by an angle amounts to rotating its polarization vector by the same angle.
We shall see in the following that such a rotation can be fairly implemented by the optical rotation \cite{Dama}, the rotation of the polarization vector of a plane wave as it travels through an optically-active medium.

\subsection{Quantum-mechanical description of optical rotation and Berry's phase}

First of all, we assume that the unit Jones wavefunction in Eqs. (\ref{a-k}) is the eigenvector of the Pauli matrix $\hat{\sigma}_1$,
$
\tilde{\alpha}_{\sigma_1}=\frac{1}{\sqrt 2}
                          \bigg(\begin{array}{c}
                                  1 \\
                                  \sigma_1
                                \end{array}
                          \bigg),
$
where $\sigma_1=\pm 1$ is the eigenvalue. In this case, the polarization vectors (\ref{a-k}) are linearly polarized,
\begin{subequations}
\begin{align}
  \mathbf{a}_{\mathbf{I},  \mathbf{k}_0, \sigma_1} &
 =\frac{1}{2} [(1+\sigma_1) \mathbf{u}_0  +i(1-\sigma_1) \mathbf{v}_0],  \label{as1} \\
  \mathbf{a}_{\mathbf{I}', \mathbf{k}_0, \sigma_1} &
 =\frac{1}{2} [(1+\sigma_1) \mathbf{u}'_0 +i(1-\sigma_1) \mathbf{v}'_0]. \label{as1'}
\end{align}
\end{subequations}
So Eq. (\ref{a'-a-k}) describes a rotation of the plane of polarization,
\begin{equation}\label{a'-a-s1}
    \mathbf{a}_{\mathbf{I}', \mathbf{k}_0, \sigma_1}
   =\exp [-i(\hat{\mathbf \Sigma} \cdot \mathbf{w}_0) \phi]
    \mathbf{a}_{\mathbf{I},  \mathbf{k}_0, \sigma_1}.
\end{equation}
Secondly, we assume that the unit Jones wavefunction is the eigenvector of the Pauli matrix $\hat{\sigma}_2$,
$
\tilde{\alpha}_{\sigma_2}=\frac{1}{\sqrt 2}\bigg(\begin{array}{c}
                                                   1 \\
                                                   i \sigma_2
                                                 \end{array}
                                           \bigg)
$,
with eigenvalue $\sigma_2= \pm 1$. The polarization vectors (\ref{a-k}) are also linearly polarized,
\begin{subequations}
\begin{align}
  \mathbf{a}_{\mathbf{I},  \mathbf{k}_0, \sigma_2} &
 =\frac{1}{\sqrt 2} e^{i \sigma_2 \frac{\pi}{4}} (\mathbf{u}_0+\sigma_2  \mathbf{v}_0),   \label{as2}\\
  \mathbf{a}_{\mathbf{I}', \mathbf{k}_0, \sigma_2} &
 =\frac{1}{\sqrt 2} e^{i \sigma_2 \frac{\pi}{4}} (\mathbf{u}'_0+\sigma_2 \mathbf{v}'_0), \label{as2'}
\end{align}
\end{subequations}
and are connected with each other by a rotation,
\begin{equation}\label{a'-a-s2}
    \mathbf{a}_{\mathbf{I}', \mathbf{k}_0, \sigma_2}
   =\exp [-i(\hat{\mathbf \Sigma} \cdot \mathbf{w}_0) \phi]
    \mathbf{a}_{\mathbf{I},  \mathbf{k}_0, \sigma_2}.
\end{equation}
Finally, we assume that the unit Jones wavefunction is the eigenvector of the helicity $\hat{\sigma}_3$,
$
 \tilde{\alpha}_{\sigma_3}=\frac{1}{2}
                           \bigg(\begin{array}{c}
                                   1+\sigma_3 \\
                                   1-\sigma_3
                                 \end{array}
                           \bigg),
$
with the eigenvalue $\sigma_3=\pm 1$. The polarization vectors (\ref{a-k}) now become circularly polarized,
\begin{eqnarray*}
  \mathbf{a}_{\mathbf{I}; \mathbf{k}_0, \sigma_3} &=&
  \frac{1}{\sqrt 2} (\mathbf{u}_0 +i \sigma_3 \mathbf{v}_0), \\
  \mathbf{a}_{\mathbf{I}';\mathbf{k}_0, \sigma_3} &=&
  \frac{1}{\sqrt 2} (\mathbf{u}'_0+i \sigma_3 \mathbf{v}'_0).
\end{eqnarray*}
In this case, the rotation operator in Eq. (\ref{a'-a-k}) reduces to a helicity-dependent phase factor,
\begin{equation}\label{a'-a-s3}
    \mathbf{a}_{\mathbf{I}'; \mathbf{k}_0, \sigma_3}
   =\exp (-i \sigma_3 \phi) \mathbf{a}_{\mathbf{I}; \mathbf{k}_0, \sigma_3},
\end{equation}
which means that for the two orthogonal eigenstates of the helicity, rotating their SV degrees of freedom by a same angle will lead to opposite phases.
Eqs. (\ref{a'-a-s1}), (\ref{a'-a-s2}), and (\ref{a'-a-s3}) describe all the characteristics of the rotation of the polarization vector of a plane wave as it travels through an optically-active medium.
It is thus concluded that the optical rotation is one of the physical processes that change the SV degree of freedom with the intrinsic quantum number remaining fixed.

It is interesting to note that the rotations (\ref{a'-a-s1}) and (\ref{a'-a-s2}) are classically indistinguishable for the polarization vectors (\ref{as1}) and (\ref{as2}) are connected with each other only by a rotation of $\frac{\pi}{4}$. But quantum-mechanically, they describe rotations of different quantum states. What is rotated in each case is the SV degree of freedom. The intrinsic quantum numbers $\sigma_1$ and $\sigma_2$ remain unchanged.
More interestingly, even though the polarization states $\mathbf{a}_{\mathbf{I}'; \mathbf{k}_0, \sigma_3}$ and $\mathbf{a}_{\mathbf{I}; \mathbf{k}_0, \sigma_3}$, being different from each other by a helicity-dependent phase, are classically indistinguishable, they are quantum-mechanically different. This result can be explained as follows.

We have seen that the helicity is not only the amplitude of the spin in the laboratory reference system but also the longitudinal component of the intrinsic degree of freedom in the LRS. We have also seen that what is canonically quantized is not the spin but the intrinsic degree of freedom.
So quantum-mechanically, the eigenvalue of the helicity should be interpreted as an intrinsic canonical variable of the photon in some LRS. Even though the polarization states $\mathbf{a}_{\mathbf{I}'; \mathbf{k}_0, \sigma_3}$ and $\mathbf{a}_{\mathbf{I}; \mathbf{k}_0, \sigma_3}$ are the eigenstates of the helicity with the same eigenvalue $\sigma_3$, their helicities as the longitudinal components of their intrinsic degrees of freedom are not specified in the same LRS. In view of this, we can say that the helicity-dependent phase in Eq. (\ref{a'-a-s3}) is the phase that Berry \cite{Berry} discovered with the help of quantum adiabatic processes.

\section{Conclusions}\label{Remarks}

In conclusion, we found by exploiting the nonholonomic characteristic of the quantum constraint (\ref{TC}) that the intrinsic degree of freedom of the photon is just a local property with respect to the LRS. It is represented in the local representation by the Pauli matrices (\ref{PM}).
The SV to fix the LRS appears as another degree of freedom with which the intrinsic degree of freedom joints in determining the polarization state of the photon in the laboratory reference system. Changing the SV degree of freedom of the eigenstate of the helicity will give rise to Berry's phase.

The difference between the Jones wavefunction and the vector wavefunction has great implications. Since the vector wavefunction is defined in the laboratory reference system, what is conjugate to the momentum via the Fourier integral of the vector wavefunction is the position of the photon in the laboratory reference system. Due to the constraint (\ref{TC}) on the vector wavefunction, the Cartesian components of the laboratory position do not commute \cite{Jord, Hawt}.
In contrast, considering that the Jones wavefunction is defined in the LRS as we explained, what is conjugate to the momentum via the Fourier integral of the Jones wavefunction cannot be the position of photon in the laboratory reference system. It should be interpreted as the position of the photon in the LRS. Because the two components of the Jones wavefunction are independent, the position in the LRS is commutative and is therefore canonically conjugate to the momentum in the sense that it satisfies the canonical commutation relations with the momentum \cite{CDG}.
The discussions on the canonical position are beyond the scope of this paper and will be presented elsewhere.


\end{document}